# Quantum Oscillations at Integer and Fractional Landau Level Indices in ZrTe$_5$

W. Yu[1], Y. Jiang[2], J. Yang[2], Z. L. Dun[3], H. D. Zhou[3], Z. Jiang[2], P. Lu[1], and W. Pan[1]

[1]Sandia National Laboratories, Albuquerque, New Mexico 87185, USA

[2]School of Physics, Georgia Institute of Technology, Atlanta, Georgia 30332, USA

[3]Department of Physics and Astronomy, University of Tennessee, Knoxville, Tennessee 37996, USA

ABSTRACT

A three-dimensional (3D) Dirac semimetal (DS) is an analogue of graphene, but with linear energy dispersion in all (three) momentum directions. 3D DSs have been a fertile playground in discovering novel quantum particles, for example Weyl fermions, in solid state systems. Many 3D DSs were theoretically predicted and experimentally confirmed. We report here the results in exfoliated ZrTe$_5$ thin flakes from the studies of aberration-corrected scanning transmission electron microscopy and low temperature magneto-transport measurements. Several unique results were observed. First, an anomalous-Hall-effect-like behavior was observed around zero magnetic field (B). Second, a non-trivial Berry's phase of π was obtained from the Landau level fan diagram of the Shubnikov-de Haas oscillations in the longitudinal resistivity $\rho_{xx}$. Third, $\rho_{xx}$ shows linear *B* field dependence in the quantum limit. Most surprisingly, quantum oscillations were observed at fractional Landau level indices N = 2/3 and 2/5, demonstrating strong electron-electron interactions effects in ZrTe$_5$.

Since the discovery of graphene[1, 2], Dirac materials have attracted tremendous attention due to their extraordinary electronic properties and great potential for applications in next generation electronic devices. Recently, research has extended to the search for three-dimensional (3D) Dirac semimetal (DS).[3-11] Unlike the two-dimensional (2D) Dirac point in graphene or on the surface of 3D topological insulators, DSs hold 3D Dirac points with linear energy dispersion along all three momentum directions. DSs have generated a great deal of current excitements and make it possible to study quantum dynamics of relativistic field theory in condensed matter systems.[10] Moreover, these topological materials are believed to be useful in future topological quantum computation. Shortly after earlier theoretical predictions, many material systems have been experimentally confirmed to be 3D DSs. For example, using angle-resolved photoemission spectroscopy (ARPES), three groups were able to observe 3D Dirac fermions in $Na_3Bi$ and $Cd_3As_2$ single crystals.[7-9, 12, 13] Questions on whether $ZrTe_5$ being a 3D DS, however, remain unsettled. In a recent theoretical study, it was shown that single-layer $ZrTe_5$ is a quantum spin Hall insulator while bulk $ZrTe_5$ is very close to the phase transition boundary between a weak and strong topological insulator.[14] On the other hand, the electronic band structure of 3D $ZrTe_5$ measured by ARPES is consistent with that expected for a 3D DS.[10] In addition to ARPES measurements, magneto-infrared spectroscopy at low temperature in high magnetic fields further shows strong evidence of inter-Landau-level transitions resulting from Dirac fermions.[11, 15] Quantum transport studies have also yielded intriguing results, such as the chiral magnetic effect.[10, 16] To date, most of the transport studies in $ZrTe_5$ were performed at relatively higher measurement temperatures, at which electron-phonon interactions can mask subtle correlation effects induced by strong electron-electron interactions. In this Letter, we show, by lowering the measurement temperature to 0.3 K, a striking observation of quantum oscillations at fractional Landau level indices.

ZrTe$_5$ has been studied in the past and is known for the resistivity anomaly and large thermoelectric power.[17, 18] It is a layered material similar to graphite. The crystal structure[14, 17, 19-21] contains chains of ZrTe$_6$ prisms running parallel to the a-axis, as shown in Figure 1a. These prismatic chains are linked along the c-axis via zigzag chains of Te atoms to form 2D planes, which stack along the b axis. In other words, ZrTe$_5$ is a layered material similar to graphite. Each layer is weakly bonded making it easy to mechanically exfoliate thin flakes of ZrTe$_5$. Figure 1c shows the projected arrangement of atoms in the a-c plane obtained from an aberration-corrected scanning transmission electron microscope (AC-STEM) taken with a high-angle annular dark-field (HAADF) detector. The high-resolution image in [010] projection in Figure 1c reveals the atomic level details of the a-c plane of ZrTe$_5$. To our knowledge, this is for the first time a high-resolution STEM HAADF image has been taken for ZrTe$_5$. The main difficulty of obtaining the STEM image is due to ZrTe$_5$ being extremely sensitive to electron radiation. From the HAADF image, we were able to determine the lattice constants of the a-c plane, and a = 0.396 nm and c = 1.382 nm (Figure 1d). The characteristic STEM pattern (i.e., an array of 5 spots with a bright one in the middle and four dim ones aside) can be explained by taking into account a half-lattice shift along the [100]-lattice direction between two adjacent layers in the [010] direction, as demonstrated in Figure 1b. The electron diffraction pattern is shown in Figure 1e. The crystal lattice constants deduced from this pattern are consistent with those obtained in Figure 1c.

In Figure 2a, we show a Hall bar device made of an exfoliated thin flake. The thickness of the flake is ~620 nm. Temperature (T) dependence of the longitudinal resistivity $\rho_{xx}$ of our ZrTe$_5$ thin flake at zero magnetic (*B*) field was measured and the details can be found in the Supporting Information Figure S1. A characteristic resistivity peak is observed at T~170 K, close to the values reported in previous work.[17, 22] Figure 2b shows the longitudinal magneto-resistivity $\rho_{xx}$ and the Hall resistivity $\rho_{xy}$ versus *B* (along *b* axis, normal to the cleavage plane) at *T* = 0.3 K. A few electronic transport features are worthwhile emphasizing here. First, $\rho_{xx}$ shows a positive, quadratic field dependence around *B* = 0. The onset of the Shubnikov-de Haas (SdH) oscillations occurs at *B* ~ 0.34 T as shown in Figure 3c. From the criterion of $\mu_q B = 1$ for the onset of SdH oscillations, a quantum mobility of $\mu_q \sim 3 \times 10^4$ cm$^2$V$^{-1}$s$^{-1}$ is obtained. The strong oscillations at *B* ~ 1.5 T and 3.0 T are still well visible at high temperatures up to 18 K. Beyond the last SdH oscillation at *B* ~ 3.0 T (corresponding to the N = 1 Landau level index), $\rho_{xx}$ assumes a linear *B* field dependence. We note here that a linear magneto-resistance has been observed for massless Dirac fermions in the quantum limit when all the carriers occupy the lowest Landau level[23-26], due to their linear energy dispersion. We believe this mechanism is also responsible for the linear *B* dependence in our sample, as we show as follows that the carriers in our sample are Dirac fermions. The Hall resistivity shows a negative slope, from which we conclude that the carriers are electrons in our device. Most surprisingly, the Hall resistivity displays an AHE-like behavior around the zero magnetic field. Similar AHE-like behavior was previously reported in Bi$_{1-x}$Sb$_x$[27], which was attributed to be a topological effect. In the high field regime, the Hall resistivity is generally linear with undulations concomitant with the SdH oscillations. The linear fit to the high field part yields a bulk carrier density of $n^{3D} = 1.8 \times 10^{18}$cm$^{-3}$. With this value, a 2D density of $n^{2D} = 1.1 \times 10^{14}$ cm$^{-2}$ is deduced, after taken into account the thickness (~ 620 nm) of the thin flake. We note that this deduced $n^{2D}$ is much higher than the electron density ( $n \sim 1.8 \times 10^{11}$ cm$^{-2}$) obtained from the SdH oscillations analysis (as we show below). To explain this discrepancy, we speculate that only the top layer of our ZrTe$_5$ thin flake participates in electron transport, similar to what has been observed in the multilayer epitaxial graphene on the c-face of SiC substrate.[28] Indeed, using the lattice constant b = 1.45 nm[17], a 2D density $n^{2D} = 2.6 \times 10^{11}$ cm$^{-2}$ is obtained, now much closer to that obtained from SdH oscillations.

To understand whether the SdH oscillations are of 2D nature, we carried out magneto-transport measurements in tilt magnetic fields. The tilt angle $\theta$ between $B$ and $b$ axis can be varied from 0 to 90° as schematically shown in Figure 2e. The angle is calibrated by measuring the quantum oscillations of an InAs quantum well that was mounted together with ZrTe$_5$ on the same chip carrier. In order to better reveal the SdH oscillations as a function of title angle, we subtract a smooth background from the measured longitudinal resistivity $\rho_{xx}$. In Figure 2c, we display the oscillatory component ($\Delta\rho_{xx}$) at T = 0.3K, obtained by subtracting the $\rho_{xx}$ curve measured at $T =$ 28 K where no SdH oscillations are observed. Well-developed quantum oscillations are seen up to the Landau level index number N = 11. Figure 2f shows the contour plot of $\Delta\rho_{xx}$ as a function of $B$ at various tilt angle $\theta$. The oscillation extrema shift to higher magnetic fields systematically with increasing $\theta$. Indeed, the magnetic field position of the N = 1 Landau level versus $\theta$ can be well fitted using $B_{N=1}(\theta = 0)/cos(\theta)$ as shown by the dashed line in Figure 2f. From this data, we conclude that the electrons that contribute to the SdH oscillations are of 2D nature.

Having confirming the 2D nature of the electron gas in our ZrTe$_5$ device, we are ready to further determine whether the 2D carriers are Dirac fermions. In so doing, we follow the well-developed methodology and construct the Landau level fan diagram. To do this, we assign a Landau level index number N (N+1/2) to each $\Delta\rho_{xx}$ minimum (maximum)[29], as shown in Figure 2c. In the whole field region, N increases by 1 between the adjacent oscillations, indicating the spin degeneracy is not lifted in our device. In the Landau fan diagram (Figure 2d), the data points fall on a straight line and the solid line represents the best linear fit. The linear extrapolation gives an intercept of -0.45[30]. The observation of a non-zero Berry's phase in our ZrTe$_5$ device indicates that the electrons responsible for the SdH oscillations are Dirac fermions. The same conclusion has also been reached in other studies, such as magneto-infrared[15], ARPES[10], and magneto-transport[11]. From the slope of the linear fit in Figure 2d a 2D carrier density of $n = 1.8 \times 10^{11}$ cm$^{-2}$ is deduced.

We now deduce the effective mass *m\** of the 2D carries from the temperature dependence of the well-developed SdH oscillations (Figure 3a). The SdH oscillation amplitude can be described by the Lifshitz-Kosevich equation[1]

$$\Delta\rho_{xx} \propto \frac{2\pi^2 k_B m^* T/\hbar eB}{\sinh(2\pi^2 k_B m^* T/\hbar eB)}. \tag{1}$$

In Eq. (1), $k_B$ is the Boltzmann's constant, $\hbar$ is the reduced Plank's constant, and *e* is the electron charge. Figure 3b shows the oscillation amplitudes as a function of temperature for different magnetic fields. The effective mass can be extracted by fitting the temperature dependence of the oscillation amplitude (solid lines in Figure 3b). In Figure 3c, *m\** is plotted as a function of magnetic field. It shows strong magnetic field dependence. In fact, it follows generally a quadratic *B* dependence, and increases from 0.022 m$_e$ at *B* ~ 0.7 T to 0.044 m$_e$ at *B* ~

1.5 T. Following this quadratic $B$ dependence, the zero $B$ field effective mass, $m^* = 0.015$ $m_e$, can be deduced. The effective mass at $B = 2.1$ T apparently deviates from this quadratic dependence, and is smaller. One possible reason for a smaller mass at this $B$ field is due to a relatively large Zeeman splitting. Indeed, the effective g-factor in ZrTe$_5$ was deduced to be around ~ 16-22[15]. At $B = 2.1$ T, the Zeeman splitting is about 22-31 K, comparable to the Landau level disorder broadening ($\Gamma$) of 30 K. Here the Landau level broadening is estimated from the quantum mobility and $\Gamma = \frac{\hbar e}{\mu_q m^*}$ ~ 30 K. The contribution from two partially spin resolved Landau levels may be responsible for a smaller effective mass at $B = 2.1$ T.

With the zero field $m^*$ and $k_F = \sqrt{2\pi n} = 1.06 \times 10^8$ m$^{-1}$, the Fermi velocity of $v_F = \frac{\hbar k_F}{m^*} = 8.2 \times 10^5$ m/s and the Fermi level $E_F = \hbar k_F v_F$ ~ 60 meV are deduced for our sample. We note here that $v_F$ obtained here is higher than those reported in Refs. 9 and 14 from magneto-infrared measurements, but comparable to that in Ref. 8 obtained in the ARPES measurements.

In the following, we present the most striking result in our experiment – the observation of quantum oscillations at fractional Landau level indices in the quantum limit regime. In Figure 4, we plot $\Delta\rho_{xx}$ (after subtracting the linear background in the high magnetic field regime) as a function of $B$ at two temperatures of 0.3 K and 10 K. At 0.3 K, additional minima at $B$ ~ 3.8, and 5.4 T are clearly visible beyond the strongest N = 1 quantum oscillation. These minima are considerably weaker than those at integer Landau level indices, and they have disappeared at $T$ ~ 10 K. We point out here that these minima are not due to experimental artifacts, as they were reproducible in different cool-downs of the same sample, and in a different sample (more details can be found in Supporting Information Figure S5). The fractional Landau indices of N = 2/3 and 2/5 can be assigned to these two minima, respectively, as shown by the dashed lines in Figure 4. There are two more weak minima at B ~ 3.4 and 4.1T, which can be assigned to N = 4/5 and 3/5, respectively. However, no minima were observed at N = 4/5 and 3/5 in another sample we examined (as shown in Supporting Information Figure S5). We notice that similar anomalous resistivity minima at fractional Landau indices were also observed in the quantum limit in a topological insulator of $Bi_2Se_3$[31]. There, they were viewed as the precursors to the fractional quantum Hall effect (FQHE). Following this same argument, we also attribute these resistivity minima with fractional Landau indices in our $ZrTe_5$ sample as the developing FQHE states. The occurrence of these quantum oscillations at fractional Landau level indices represents compelling evidence of electron-electron interactions induced many-body correlation effects in $ZrTe_5$.

In summary, we have systematically studied $ZrTe_5$, a new candidate of three-dimensional Dirac semimetals. High resolution HAARD image was taken for the first time and the obtained crystal lattice constants for the a-c plane are consistent with previous reports. Magneto-transport measurements were carried out in $ZrTe_5$ thin flake. Shubnikov-de Haas Oscillations were observed. A non-trivial $\pi$ Berry's phase is obtained in the Landau fan diagram, manifesting the nature of Dirac semimetal phase in $ZrTe_5$. Angular-dependent measurements reveal the 2D-like behavior of Dirac fermions. Like graphene, the unique layer structure and ultra-high mobility make $ZrTe_5$ a promising material system for future high speed electronic devices.

**Methods.** *ZrTe$_5$ synthesis.* ZrTe$_5$ polycrystalline sample was prepared by reacting appropriate ratio of Zr and Te in a vacuumed quartz tube at 450 degree for one week. The ZrTe$_5$ single crystal sample was prepared by chemical vapor transport technique[32, 33]. The transport agent is iodine and the transport temperature is from 530 degree to 450 degree. The transport time is around 20 days.

*Device fabrication.* By exfoliating ZrTe$_5$ single crystal, flakes were transferred onto 1-µm-thick silicon dioxide on P-Si <100> substrates. 300-nm-thick Pd electrodes were defined using e-beam lithography technique followed by physical vapor deposition. A standard lift-off process is employed.

*Characterization.* The thickness of thin ZrTe$_5$ flake was measured using atomic force microscope (Veeco D3100 with a Nanoscope IVA Controller).

*Transport measurement.* The measurement was performed in a He-3 cryostat equipped with a superconducting magnet. The magnetic field can go up to 7.5 T. A standard low-frequency technique is used to measure the resistance with an *ac* excitation current of 5 µA. Angular-dependent measurement was carried out by mounting the sample on a rotary stage. ZrTe$_5$ and InAs were put side by side on the same chip carrier. The tilt angle was calibrated by measuring the quantum states of InAs.

ASSOCIATED CONTENT

**Supporting Information**

Temperature dependence of the longitudinal resistivity, Hall resistivity at T = 0.3 K and T = 10 K, angular-dependent measurement at T = 0.3 K, SdH oscillation amplitude, Measurement results of another device. This material is available free of charge via the Internet at http://pubs.acs.org.

## AUTHOR INFORMATION

**Corresponding Author**

*Email: wnyu@sandia.gov

**Notes**

The authors declare no competing financial interests.

## ACKNOWLEDGMENT

This work was supported by the U.S. Department of Energy, Office of Science, Basic Energy Sciences, Materials Sciences and Engineering Division. Sandia National Laboratories is a multi-program laboratory managed and operated by Sandia Corporation, a wholly owned subsidiary of Lockheed Martin Corporation, for the U.S. Department of Energy's National Nuclear Security Administration under contract DE-AC04-94AL85000. Work at Georgia Institute of Technology was, in part, supported by the Department of Energy (Grant No. DE-FG02-07ER46451). $ZrTe_5$ single crystal growth at University of Tennessee was supported by the National Science Foundation (Grant No. DMR-1350002).

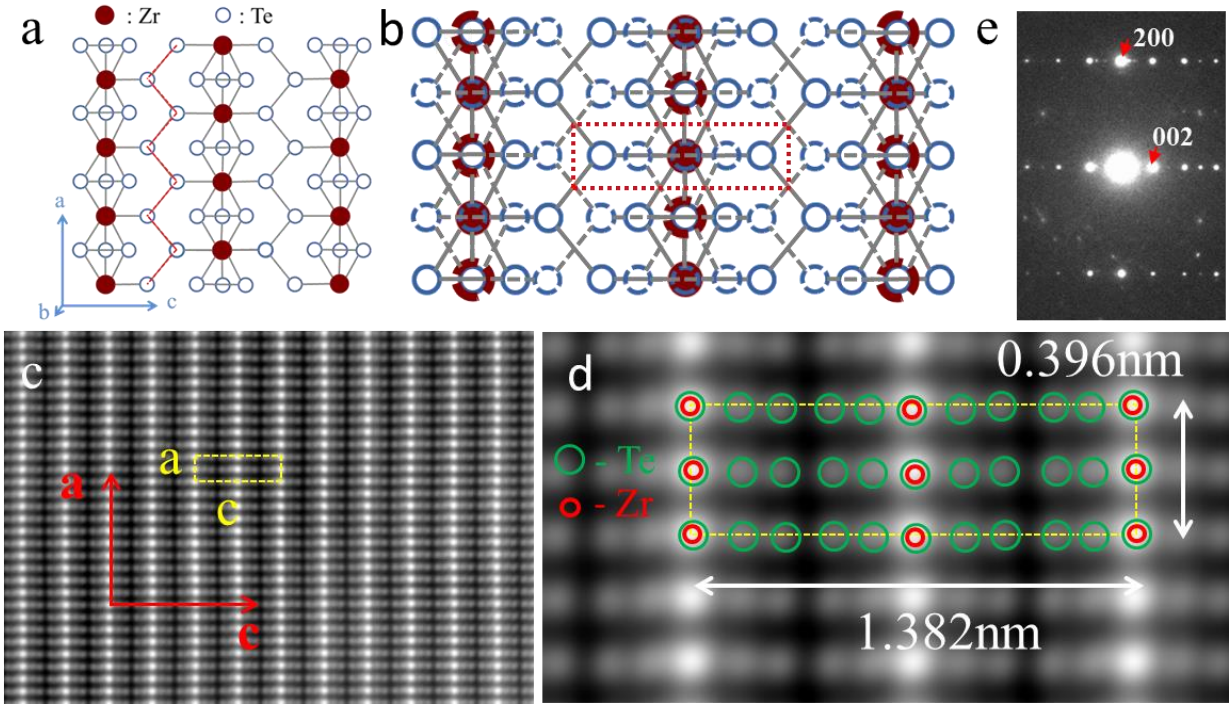

**Figure 1:** $ZrTe_5$ crystal structure. (a) Schematic top view of a single-layer structure. The dashed red line shows the zigzag chains of Te (adapted from Ref. [17]). (b) Schematic top view of a two-layer structure. The solid patterns represent the top layer and the dashed the bottom layer. The bottom layer is shifted by a half lattice-constant along the [100]-crystal direction. Due to this shift, one Zr atom overlaps with a Te atom in the projection. (c) STEM high-angle annular dark-field (HAADF) image of $ZrTe_5$ in [010] direction. (d) Zoomed-in image of the unit cell (dashed yellow rectangular) in (c). Red and green circles represent Zr and Te atoms, respectively. The bright spots are due to the overlapping of Zr and Te atoms in the projection. The measured lattice constant c = 1.382 nm and a = 0.396 nm. (e) Corresponding electron diffraction pattern in [010] direction with (200) and (002) planes as marked.

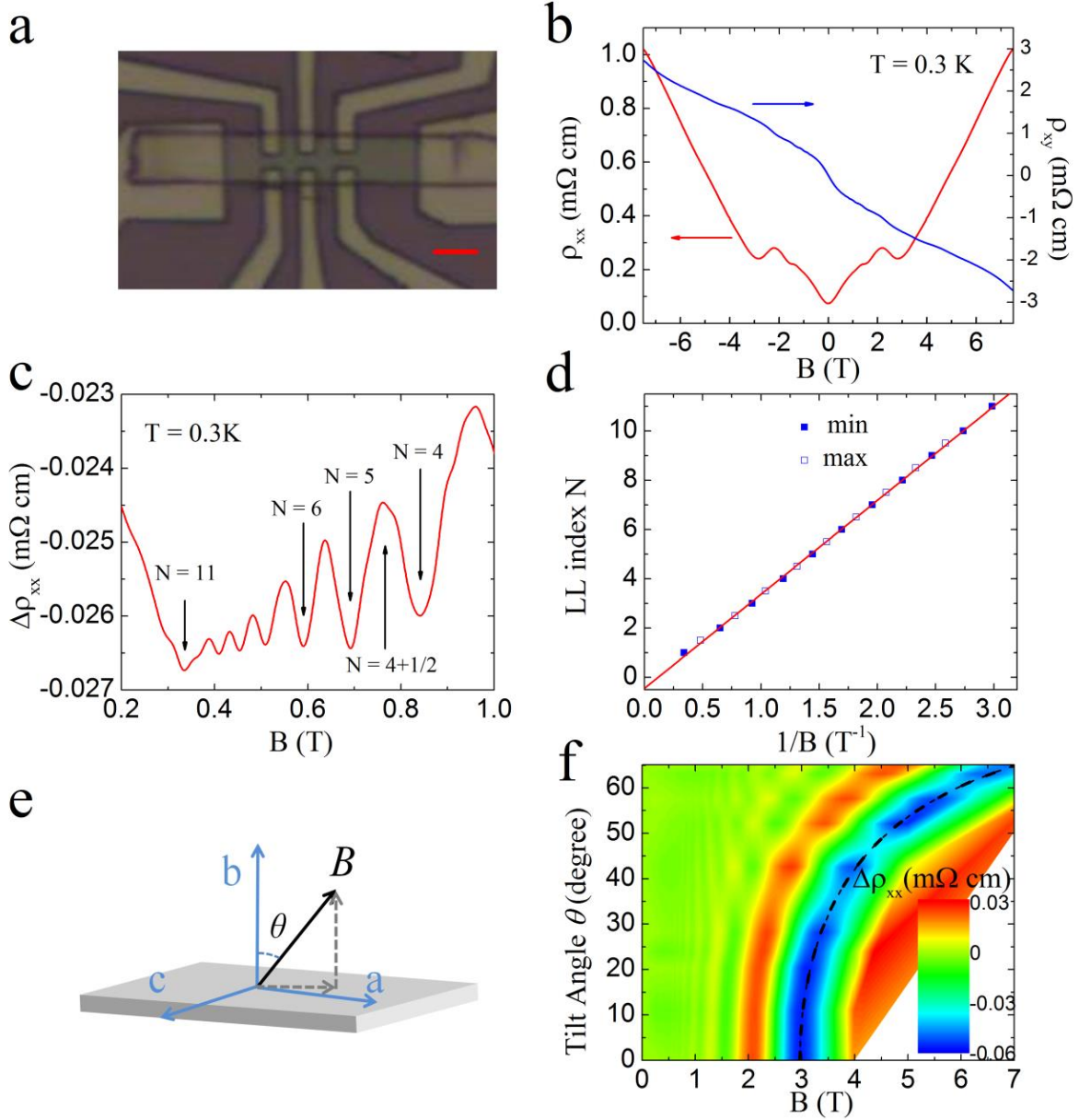

**Figure 2:** SdH oscillations in the longitudinal resistivity and tilt magnetic field dependence. (a) Optical image of the Hall bar device. The scale bar is 5 μm. (b) The longitudinal resistivity (red line) and the transverse Hall resistivity (blue line) as a function of magnetic (*B*) field at *T* = 0.3 K. *B* is along *b* axis, perpendicular to the cleavage plane. Well-developed SdH oscillations are observed. (c) *Δρ*$_{xx}$ as a function of magnetic field for tilt angle *θ* = 0°. *Δρ*$_{xx}$ is obtained by subtracting a smooth background based on the curve at $T$ = 28 K. Several LLs are labeled by the arrows. LL N = 11 corresponds to $B \sim$ 0.34 T. (d) LL index N versus 1/$B$. The closed squares represent the integer index ($\Delta\rho_{xx}$ minima) and the open squares denote the half integer index ($\Delta\rho_{xx}$ maxima). The solid line shows the best liner fit. An intercept of -0.45 is obtained, indicative of a $\pi$ Berry's phase. (e) Schematic of the sample in tilt magnetic field. (f) Contour plot of $\Delta\rho_{xx}$ with respect to tilt angle $\theta$ and magnetic field $B$. Blue color highlights the evolution of LL N = 1. The dashed line represents the best fit by $B_{N=1}(\theta = 0°)/\cos(\theta)$, indicating the 2D-like transport behavior of charged carriers.

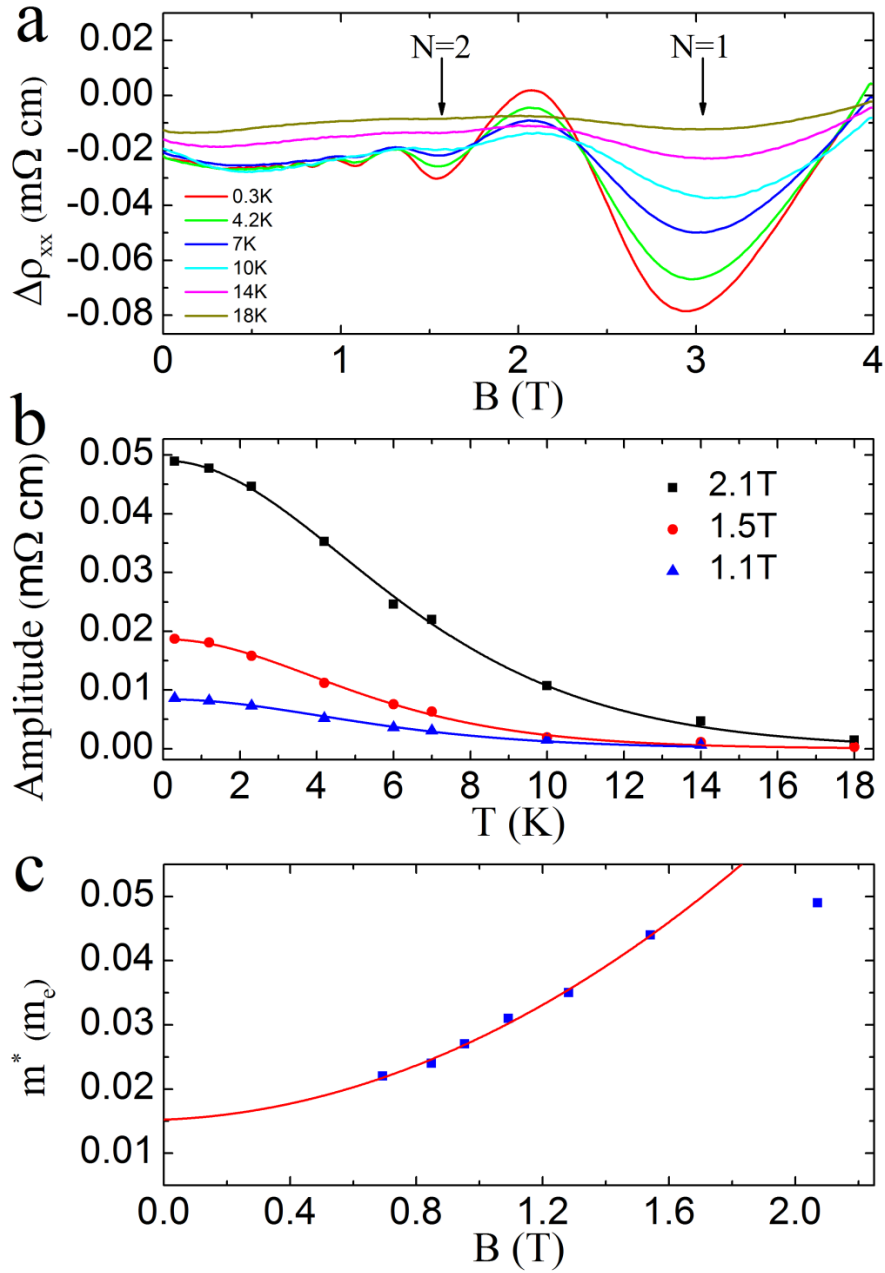

**Figure 3:** Determining effective mass from temperature dependence of SdH oscillations. (a) $\Delta\rho_{xx}$ as a function of magnetic field at different temperatures obtained by subtracting a smooth background based on the curve at $T = 28$ K. LLs $N = 1, 2$ are labeled by the arrows. (b) SdH oscillation amplitude as a function of temperature at different magnetic fields. The solid lines represent the best fit using Eq. (1).

Data points obtained at low fields can be found in the Supporting Information. (c) Extracted effective mass at different magnetic fields. The line is a quadratic magnetic field dependent fit.

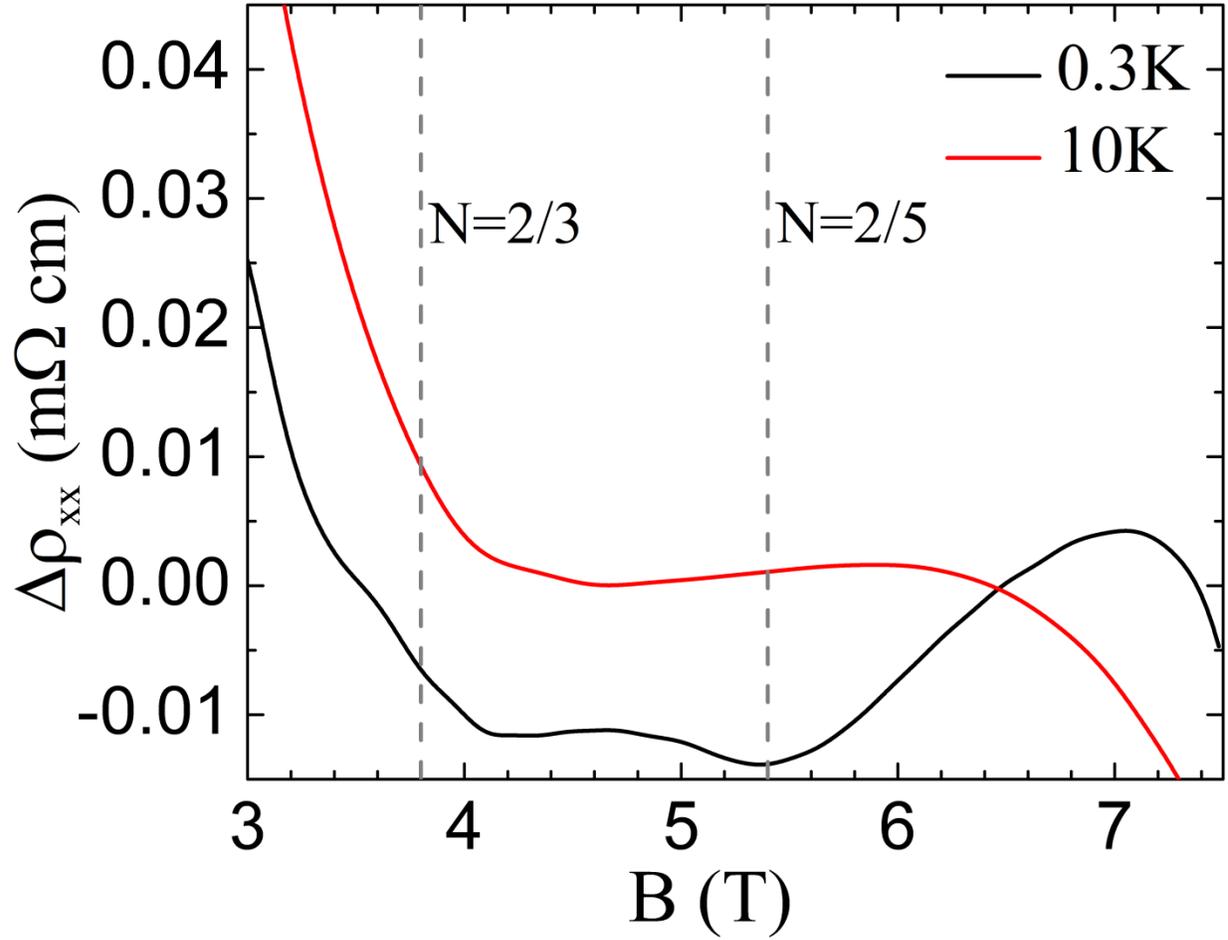

**Figure 4:** Quantum oscillations at fractional Landau level indices. $\Delta\rho_{xx}$ in the quantum limit regime is plotted as a function of magnetic field at two temperatures of 0.3K and 10 K. Here, $\Delta\rho_{xx}$ is obtained by subtracting a linear background. Fractional Landau level indices at N = 2/3 and 2/5 are labeled by the dashed lines.

# Supporting Information

## Quantum Oscillations at Integer and Fractional Landau Level Indices in ZrTe$_5$

W. Yu[1,*], Y. Jiang[2], J. Yang[2], Z. L. Dun[3], H. D. Zhou[3], Z. Jiang[2], P. Lu[1], and W. Pan[1]

*1Sandia National Laboratories, Albuquerque, New Mexico 87185, USA*

*2School of Physics, Georgia Institute of Technology, Atlanta, Georgia 30332, USA*

*3Department of Physics and Astronomy, University of Tennessee, Knoxville, Tennessee 37996, USA*

*Corresponding author: wnyu@sandia.gov

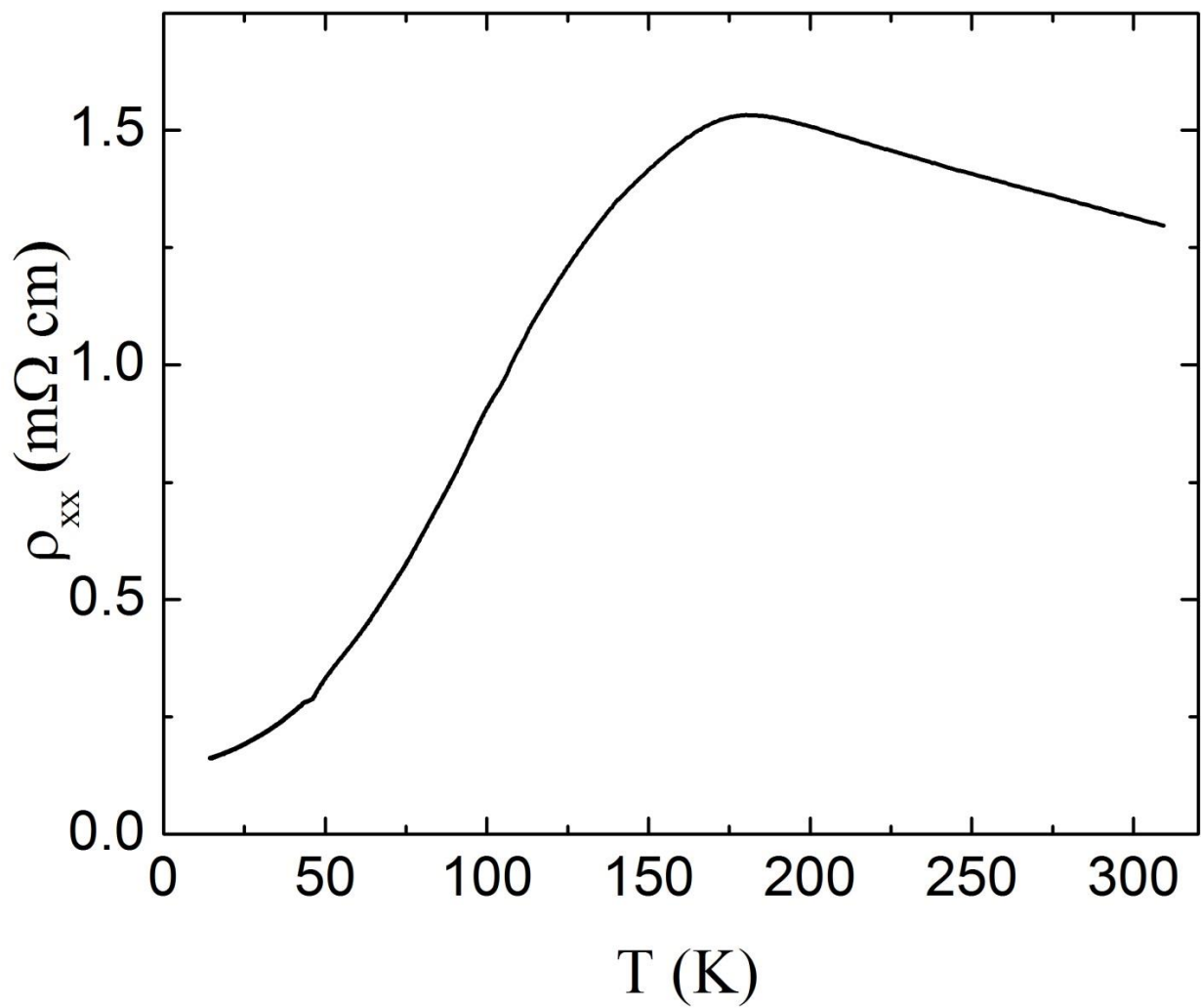

**Figure S1.** The temperature dependence of the longitudinal resistivity $\rho_{xx}$ of ZrTe$_5$ thin flake at zero magnetic field. A characteristic resistivity peak is observed at T~170 K, close to the values reported in previous work[1,2].

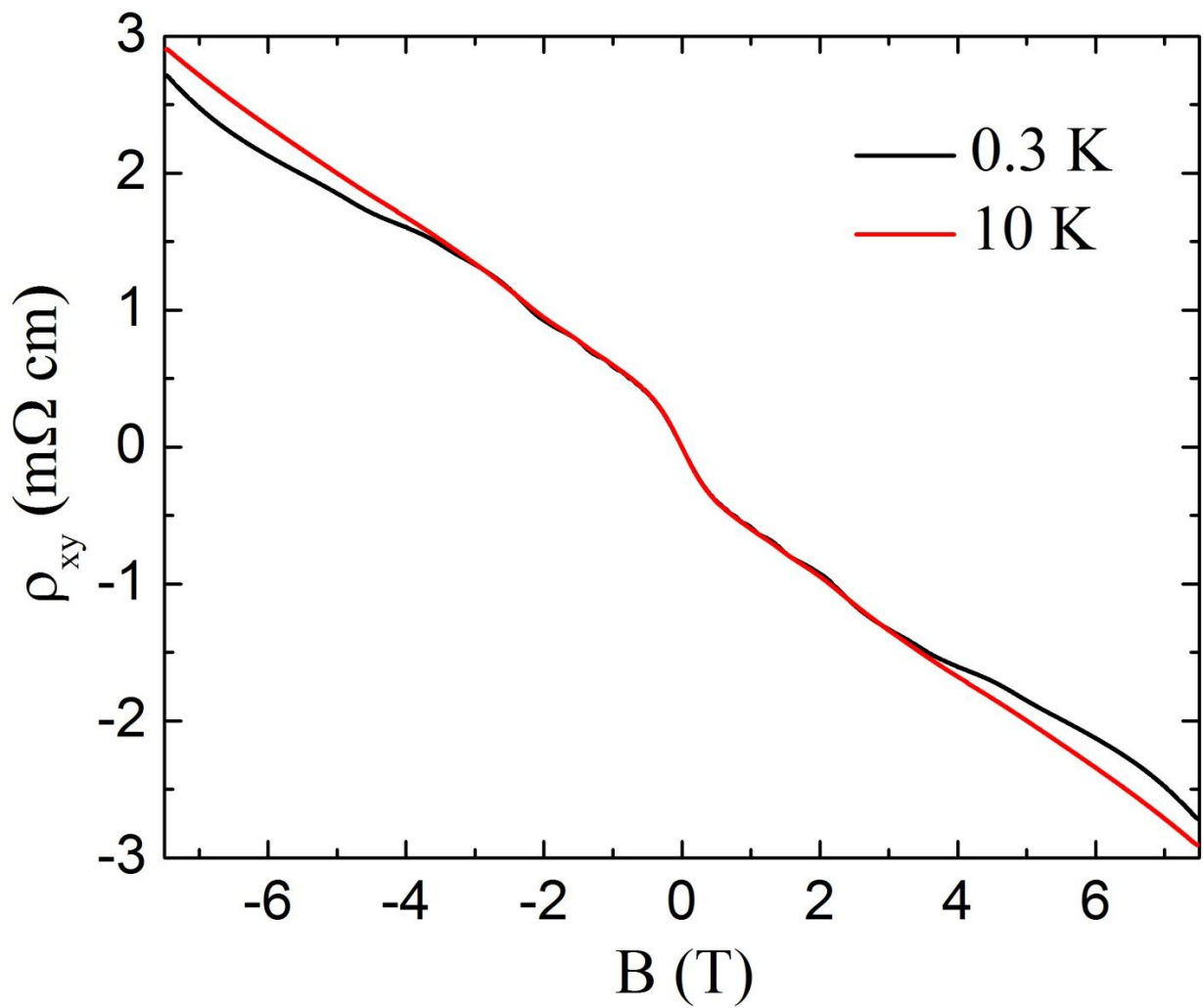

**Figure S2.** Hall resistivity at $T = 0.3$ K and $T = 10$ K, respectively. The two traces overlap with each other indicating constant carrier densities in this large temperature range.

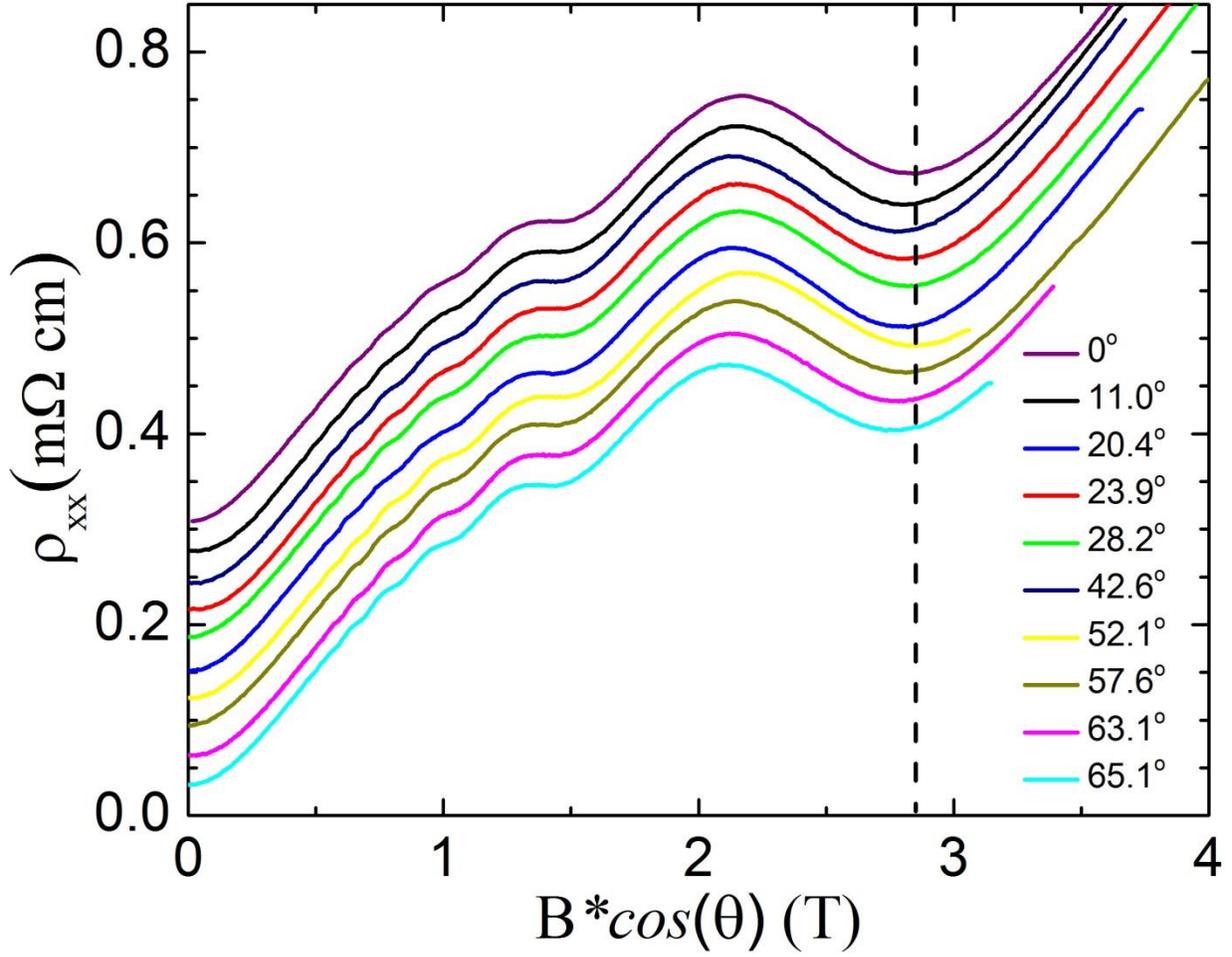

**Figure S3**. Angular-dependent measurement at $T$ = 0.3 K. $\rho_{xx}$ is plotted against $B_\perp = B\cos(\theta)$. The dashed line indicates Landau level N = 1. Curves are shifted for clarity. From this data, we can conclude that the electrons that contribute to the SdH oscillations are of 2D nature.

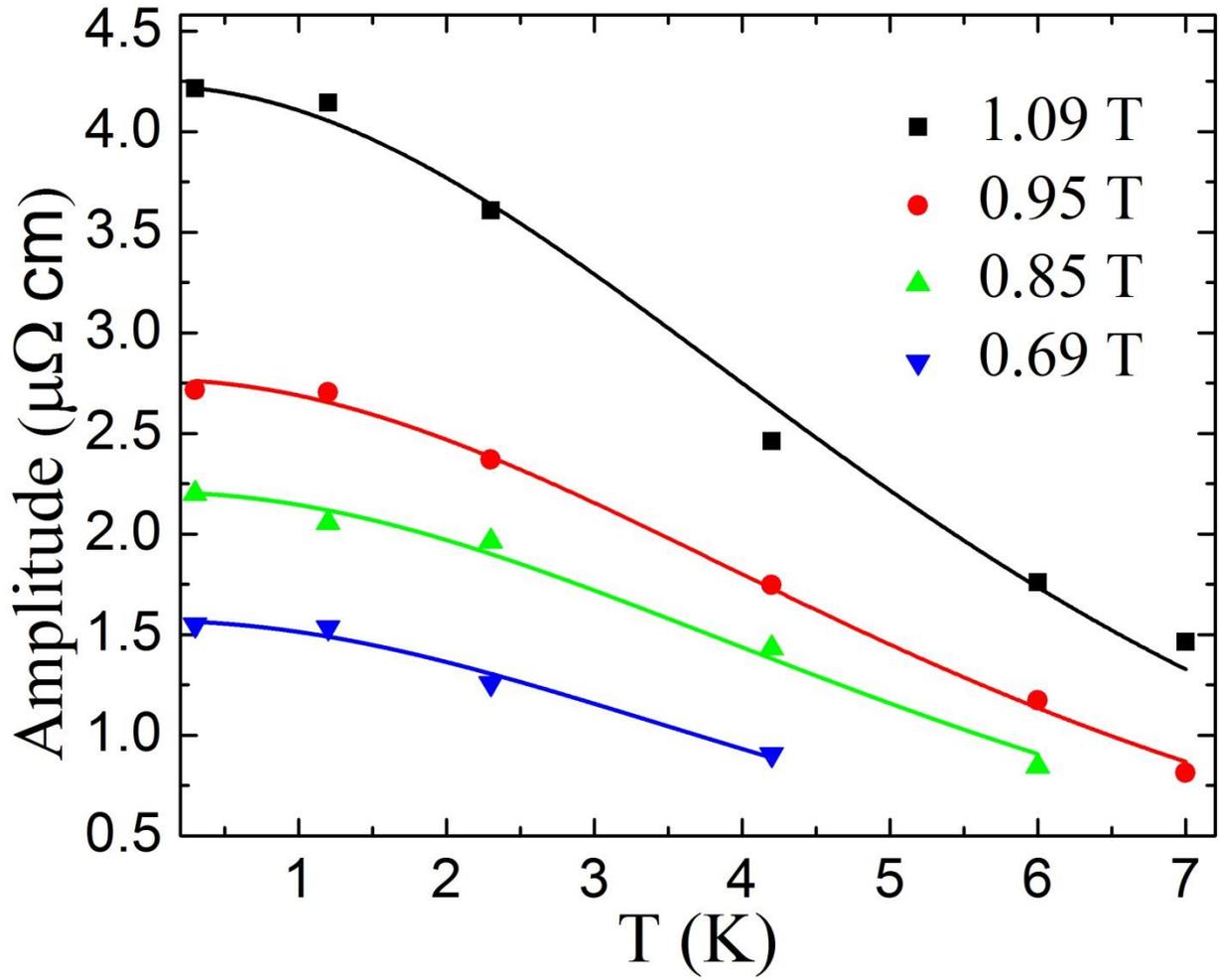

**Figure S4.** SdH oscillation amplitude as a function of temperature in the low magnetic field regime. The solid lines represent the best fit using Eq. (1).

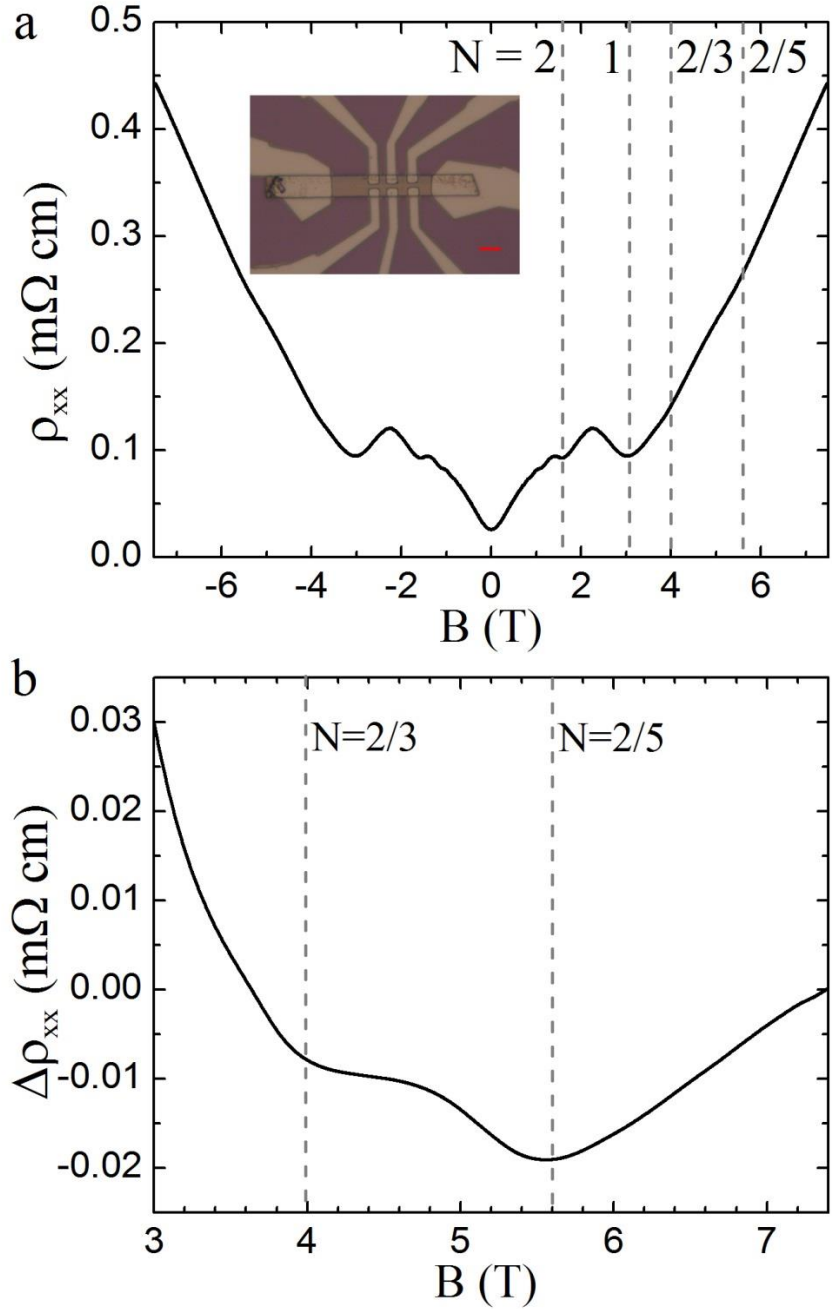

**Figure S5.** Measurement results of another device. (a) Longitudinal resistivity $\rho_{xx}$ as a function of magnetic field at $T = 0.3$ K. Landau levels are marked by dashed lines. Inset presents the optical image of the device. The thickness of ZrTe$_5$ flake is ~ 1.35 μm. The scale bar is 10 μm. (b) In the high magnetic field regime, $\Delta\rho_{xx}$ is obtained by subtracting a linear background. Fractional Landau level indices at N = 2/3 and 2/5 are labeled by the dashed lines.